\documentclass[aps,superscriptaddress,12pt]{revtex4}
\usepackage{amsfonts}
\usepackage{amsmath}
\usepackage{amssymb}
\usepackage{graphicx}

\begin{document}

\title{$N$-photon Correlation Functions in $N$-slit Diffraction Experiments with Thermal Light}
\author{Su-Heng Zhang}
\affiliation{College of Physics Science $\&$ Technology, Hebei University, Baoding
071002, China}
\author{De-Zhong Cao}
\email{dzcao@ytu.edu.cn}
\affiliation{Department of Physics, Yantai University, Yantai 264005, China}

\begin{abstract}
We discuss the analytic presentations of the high-order correlation
functions in the $N$-slit diffraction with thermal light in a recent
paper [Phys. Rev. Lett. \textbf{109}, 233603 (2012)]. Our analysis
shows that the superresolving fringes in the high-order
correlation measurement have two classical counterparts.
\end{abstract}

\maketitle

In a recent paper, Oppel \textit{et al}. \cite{oppel} obtained
superresolving fringes in an $N$-slit diffraction experiment with
thermal light sources. The analytic results of the high-order
correlation functions, with order up to 5, were given. However, the
general presentations of arbitrary $N$th-order correlation functions
with thermal light were not presented. In this paper, we
theoretically reconsider the N-slit diffraction experiment, and calculate
out the analytic results of the arbitrary Nth-order correlation functions.

The experimental setup of $N$-photon superresolving interference with
thermal light is shown in Fig. 1(a). A uniform thermal light beam passes
through $N$ slits of separation $d$, thereafter is registered by $N$
well-arranged detectors in the far-field diffraction plane. The slits are
positioned at $(2n-N-1)d/2,\;n=1,2,\cdots ,N$, i.e., all the $N$ slits are
symmetrical with respective to the origin. Among the $N$ detectors, the
first detector scans while the other $N-1$ ones are fixed at certain magic
positions, given in terms of phases $\delta _{j}=2\pi d\sin \theta
_{j}/\lambda =(j-1)2\pi /(N-1)$,$\;(j=2,\cdots ,N)$ \cite{thiel}, where
$\lambda $ is the light wavelength, and $\theta _{j}$\ is the diffraction
angle. The joint measurement of the $N$ detectors gives the $N$th-order
intensity correlation function
\begin{align}
G^{(N)}(\delta ,\delta _{2},\cdots ,\delta _{N})& =\left\langle I(\delta
)I(\delta _{2})\cdots I(\delta _{N})\right\rangle  \notag \\
& =\left\langle E^{\ast }(\delta )E^{\ast }(\delta _{2})\cdots E^{\ast
}(\delta _{N})\times E(\delta _{N})\cdots E(\delta _{2})E(\delta
)\right\rangle ,  \label{1}
\end{align}%
where $\delta =2\pi d\sin \theta /\lambda $ indicates the position of the
scanning detector.

\begin{figure}[tbp]
\includegraphics[width=5cm,angle=-90]{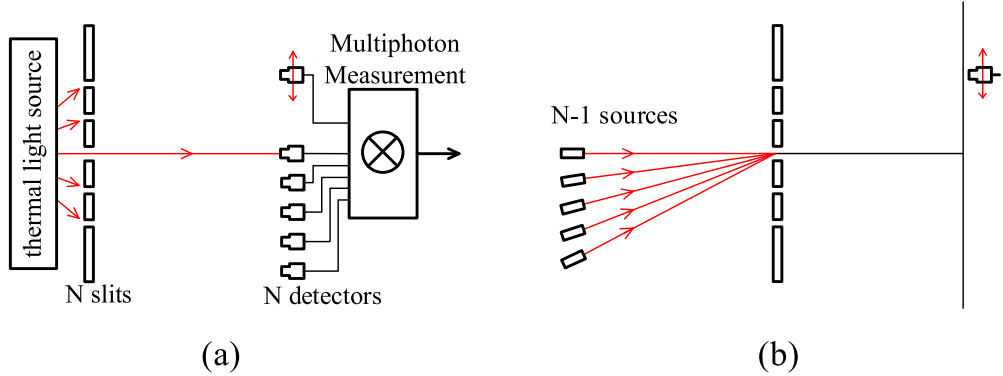}
\caption{(a) Schematic diagram for $N$-photon superresolving interference
with thermal light. One detector scans, while the other $N-1$ detectors are
fixed; (b) Unfolded setup that $N-1$ fixed detectors were regarded as light
sources. The interference fringes are registered by the moving detector. }
\end{figure}

According to the Gaussian moment theorem, one can expand the $N$-th-order
correlation function into terms of first-order correlation functions that
\cite{mandel,cao1}%
\begin{equation}
G^{(N)}(\delta ,\delta _{2},\cdots ,\delta _{N})=\sum_{N!}\mathcal{P}%
\left\langle E^{\ast }(\delta )\underline{E(\delta )}\right\rangle
\left\langle E^{\ast }(\delta _{2})\underline{E(\delta _{2})}\right\rangle
\cdots \left\langle E^{\ast }(\delta _{N-1})\underline{E(\delta _{N-1})}%
\right\rangle ,  \label{2}
\end{equation}%
where $\mathcal{P}$ represents the $N!$ possible permutations of the
underlined fields.

We assume the thermal source satisfies completely spatial incoherence. In
the far field diffraction by considering the magic positions, the
first-order correlation functions of the detected fields in Eq. (\ref{2})
are calculated out \cite{cao2}, and the results can be written as
\begin{equation}
\left\langle E^{\ast }(\delta )E(\delta )\right\rangle =I_{0}N,  \label{3}
\end{equation}%
\begin{equation}
\left\langle E^{\ast }(\delta _{j})E(\delta _{k})\right\rangle =%
\begin{cases}
I_{0}N, & \quad j=k \\
I_{0}(-1)^{j-k}, & \quad j\neq k%
\end{cases}
\label{4}
\end{equation}%
\begin{equation}
\left\langle E^{\ast }(\delta )E(\delta _{j})\right\rangle =I_{0}\frac{\sin %
\left[ \frac{N}{2}(x-2\pi \frac{j-2}{N-1})\right] }{\sin \left[ \frac{1}{2}%
(x-2\pi \frac{j-2}{N-1})\right] },  \label{5}
\end{equation}%
where $j,k=2,\cdots ,N$. Note that $\left\langle E^{\ast }(\delta )E(\delta
_{j})\right\rangle $ in Eq. (\ref{5}) represents the Fourier transform of the
$N$-slit. The $N $th-order intensity correlation function in Eq. (\ref{2}) can
be normalized as
\begin{equation}
g^{(N)}(\delta ,\delta _{2},\cdots ,\delta _{N})=\frac{G^{(N)}(\delta
,\delta _{2},\cdots ,\delta _{N})}{\left\langle E^{\ast }(\delta )E(\delta
)\right\rangle \left\langle E^{\ast }(\delta _{2})E(\delta
_{2})\right\rangle \cdots \left\langle E^{\ast }(\delta _{N})E(\delta
_{N})\right\rangle }.  \label{2a}
\end{equation}%
The denominator in Eq. (\ref{2a}) is
\begin{equation}
\left\langle E^{\ast }(\delta )E(\delta )\right\rangle \left\langle E^{\ast
}(\delta _{2})E(\delta _{2})\right\rangle \cdots \left\langle E^{\ast
}(\delta _{N})E(\delta _{N})\right\rangle =(I_{0}N)^{N},  \label{2aa}
\end{equation}%
in which the results of the first-order correlation functions (\ref{3}, \ref{4}%
) are taken into account. By categorizing all the $N!$ possible permutations
of the first-order correlation functions, we arrive at%
\begin{equation}
g^{(2)}(\delta ,\delta _{2})=1+\left\vert g^{(1)}(\delta ,\delta
_{2})\right\vert ^{2}  \label{2aaa}
\end{equation}%
for $N$=2, and
\begin{equation}
g^{(N)}(\delta ,\delta _{2},\cdots ,\delta _{N})=A+B\left\vert
\sum_{j=2}^{N}(-1)^{j}g^{(1)}(\delta ,\delta _{j})\right\vert
^{2}+C\sum_{j=2}^{N}\left\vert g^{(1)}(\delta ,\delta _{j})\right\vert ^{2},
\label{6}
\end{equation}%
for $N\geq 3$. The normalized first-order correlation function is
\begin{equation}
g^{(1)}(\delta ,\delta _{j})=\frac{\left\langle E^{\ast }(\delta )E(\delta
_{j})\right\rangle }{\sqrt{\left\langle E^{\ast }(\delta )E(\delta
)\right\rangle \left\langle E^{\ast }(\delta _{j})E(\delta
_{j})\right\rangle }}.
\end{equation}%
The three coefficients are%
\begin{align}
A& =ND_{N}(N-1),  \label{7} \\
B& =\sum_{n=0}^{N-3}D_{N}(N-3-n)\frac{(N-3)!}{(N-3-n)!},  \label{8} \\
C& =D_{N}(N-2)-B,  \label{9}
\end{align}%
where
\begin{equation}
D_{N}(M)=\sum_{n=0}^{M}\frac{N^{M-n-N}M!}{(M-n)!}\sum_{m=0}^{n}\frac{(-1)^{m}%
}{m!}.  \label{10}
\end{equation}

Let us analyze the three terms in the right side of Eq. (\ref{6}). The first
term contributes a homogeneous background to the $N$-photon joint
measurement. The last two terms represent the superresolving interference
fringes. We note that the last two terms are not the Fourier transform of
the source profile (\ref{5}) when $N>2$. By considering Eq. (\ref{5}), we
obtain the summations in Eq. (\ref{6})
\begin{equation}
\left\vert \sum_{j=2}^{N}(-1)^{j}g^{(1)}(\delta ,\delta _{j})\right\vert
^{2}=2(N-1)^{2}\left[ 1+\cos (N-1)\delta \right] ,  \label{11}
\end{equation}%
\begin{equation}
\sum_{j=2}^{N}\left\vert g^{(1)}(\delta ,\delta _{j})\right\vert ^{2}=N(N-1)
\left[ 1+\frac{2}{N}\cos (N-1)\delta \right] ,  \label{12}
\end{equation}%
both showing cosinusoidal spatial modulations.

Since the source plane can be regarded as a conjugate mirror in high-order
intensity correlation function of thermal light \cite{cao3}, we unfold the
experimental setup as shown in Fig. 1 (b). The thermal light correlation in
the experiment can become comprehensible that as if the $N-1$ fixed
detectors played the role of optical sources. Imagine the light came from
the $N-1$ fixed detectors, gone back to the thermal source plane, and gone
through the $N$-slit, and finally registered by the moving detector.
Therefore, the results in Eqs. (\ref{11}, \ref{12}) can be obtained in
single-photon diffraction of $N$-slit with coherent (\ref{11}) and incoherent
(\ref{12}) illuminations, respectively.

\end{document}